\documentclass[aps,prd,twocolumn,longbibliography,nofootinbib,showkeys]{revtex4-2}

\usepackage[english]{babel}
\usepackage[latin1]{inputenc}
\usepackage{enumerate}
\usepackage{amsmath}
\usepackage{amsthm}
\usepackage{amssymb}
\usepackage{graphicx}
\usepackage{floatflt}
\usepackage{fancybox}
\usepackage{appendix}
\usepackage{enumitem}
\usepackage{subfigure}
\usepackage{array}

\usepackage{hyperref}

\usepackage{color}

\renewcommand{\Re}{\textrm{Re}}

\begin{document}
\selectlanguage{english}

\title{Topological phases and scattering protected by projective chiral symmetry}

\author{Théo Torres}
\email{theo.torres-vicente@cnrs.fr}
\affiliation{Aix Marseille Univ., CNRS, Centrale Marseille, LMA UMR 7031, Marseille, France}
\affiliation{Université Bourgogne Europe, CNRS, Laboratoire ICB UMR 6303, 21000 Dijon, France}

\author{Renaud C\^ ote} 
\email{cote@lma.cnrs-mrs.fr}
\affiliation{Aix Marseille Univ., CNRS, Centrale Marseille, LMA UMR 7031, Marseille, France}

\author{Marc Pachebat} 
\email{pachebat@lma.cnrs-mrs.fr}
\affiliation{Aix Marseille Univ., CNRS, Centrale Marseille, LMA UMR 7031, Marseille, France}

\author{Antonin Coutant}
\email{coutant@lma.cnrs-mrs.fr}
\affiliation{Aix Marseille Univ., CNRS, Centrale Marseille, LMA UMR 7031, Marseille, France}

\date{\today}

\begin{abstract}
	The bulk-boundary correspondence states that topological properties of a bulk material guarantee the existence of protected edge modes, oftentimes associated with the attractive property of backscattering immunity. In this Letter, we consider a time reversal invariant classical system with both translation and chiral, or sublattice, symmetries. We show that when these two symmetries anti-commute, the material exhibits a topological phase protected by the same invariant as type D topological superconductors. We realize such a system experimentally and observe the existence of the associated edge modes in an acoustic network. Our measurements are in excellent agreement with theoretical predictions from discrete lattice models. Finally, we show that these edge modes are immune to backscattering for perturbations consisting in two subsequent identical defects. 
\end{abstract}


\maketitle



%
%

\medskip
\noindent
\textit{Introduction.} Topology in condensed matter physics provides a way to classify materials according to global properties of their eigenstates that remain robust against smooth deformations~\cite{Ryu2010,Teo2010}. This concept was developed after the discovery of topological insulators: a material with the bulk properties of an insulator but that can give rise to robust conducting states at their boundaries whose existence are protected by topological numbers.

The concept of topological phases is fundamentally a wave phenomenon, and hence, it can naturally go beyond the realm of condensed matter to enter various classical wave systems such as acoustics, photonics or mechanics~\cite{Huber2016,Ma2019,Price2022}. 
While topology guarantees the existence of protected edge modes, it is oftentimes associated with the attractive property of backscattering immunity. In classical wave settings however, backscattering is usually nonzero~\cite{Orazbayev2019,Deng2019,Arora2021,Leveque2023,Rosiek2023,Rechtsman2023}, and instances of perfect transmission are very seldom~\cite{Coutant2024}.  

In general, topological phases are directly linked to the presence, or absence, of extra symmetry in the system. In this work, we consider time reversal invariant lattice models with both translation symmetry and chiral, also known as sublattice, symmetry. Usually, one considers lattices where those two symmetries commute with each other, which means that the sublattice structure is invariant by translation. This is the case for the well-known Su-Schieffer-Heeger (SSH) model~\cite{Su1979,Asboth2016}. However, there is another natural class of systems, where translation swaps the role of the two sublattices~\cite{Xiao2024}. This corresponds algebraically to an anti-commutation relation between the chiral operation and a translation. This is quite similar to systems implementing projective symmetries, where spatial symmetries commutation relations pick up additional phase shifts~\cite{Chen2023,Chen2022,Xue2022,Li2022,Li2023,Jiang2024}. Following the analogy, we coin our class of systems as ``{\it projective chiral}''. We analyze here how it combines with time reversal invariance, a particularly strong constraint in classical wave systems~\cite{Caloz2018,Gan2021}. We show that this combination of symmetries gives rise to enriched band structures and scattering properties. In particular, for two dimensional lattices, it leads to a novel type of topological edge modes, protected by the same invariant as type $D$ topological superconductors~\cite{Ryu2010,Chiu2016}, whose emblematic model is given by the Kitaev chain~\cite{Kitaev2001,Alicea2012}. In addition we show that the topologically protected modes are also immune to backscattering to a certain class of defect, namely defects which are composed of two copies of arbitrary deformations.

We realise this unique topological phase experimentally by considering an acoustic system consisting of a network of cylindrical tubes. We measure and show the existence of the topologically protected edge mode, whose profile and energy are in excellent agreement with our theoretical lattice model.

%
%

\medskip
\noindent
\textit{-- Projective chiral symmetry.} We consider a class of time reversal symmetric lattice models that display chiral, i.e. sublattice symmetry. 
This means that the lattice is bipartite and can be split into two sublattices $\mathcal A$ and $\mathcal B$, with only inter-sublattice couplings. Such a model is described by a tight-binding Hamiltonian $\mathcal H = \sum t_{j j'} a_j^\dagger a_{j'}$ with $a_j$ (resp. $a_j^\dagger$) the annihilation (resp. creation) operator on site $j$. The sublattice symmetry implies that $\mathcal H$ anti-commutes with the chiral operator $\Gamma = \mathrm{diag}(-\mathbf{1}_{\mathcal A},\mathbf{1}_{\mathcal B})$. The key hypothesis in this work is that the chiral operator $\Gamma$ anti-commutes with the translation $\tau_x$ along one direction, that is: 
\begin{equation}\label{eq:PrChiral_x}
	\{\Gamma, \tau_x\} = 0 . 
\end{equation} 
In other words, a minimal unit cell does not preserve chiral symmetry because the role of sublattices $\mathcal A$ and $\mathcal B$ are swapped when moving from a unit cell to the next one. We refer to this relation~\eqref{eq:PrChiral_x} as projective chiral symmetry.

\begin{figure*}[htp]
	\centering
	\includegraphics[width=\textwidth]{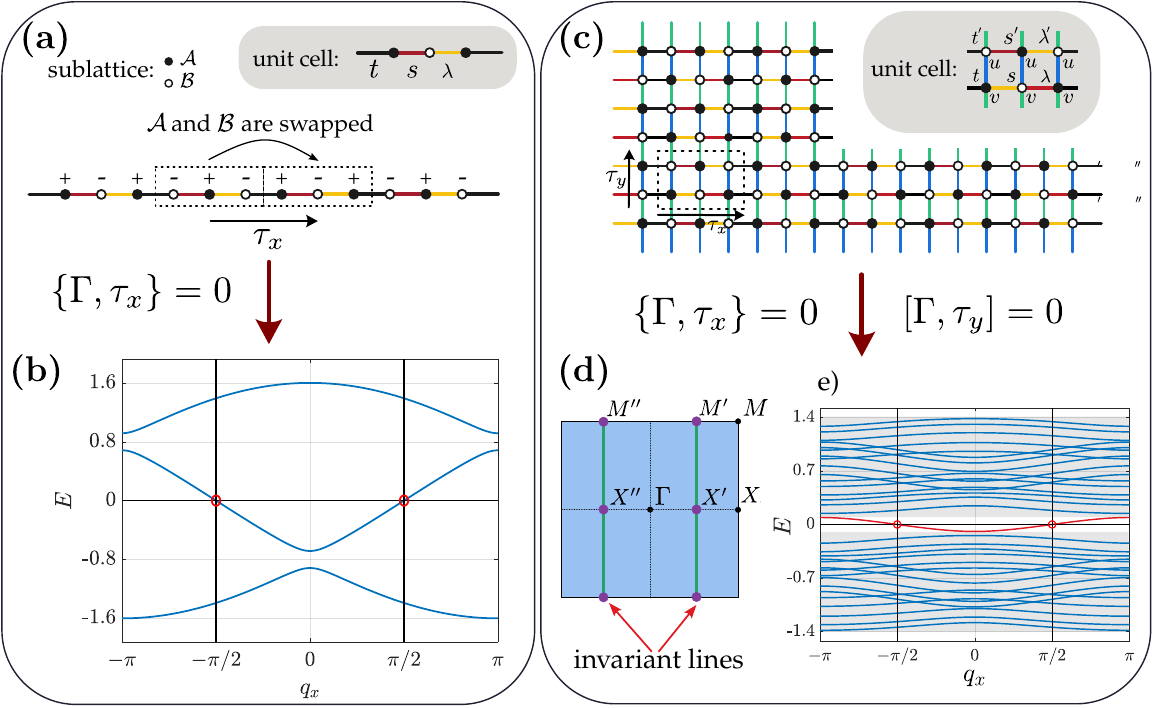} 
	\caption{(a) Lattice of the SSH3 model. The $\pm$ sign above a site indicates its sublattice ($-$ for $\mathcal A$ and $+$ for $\mathcal B$). (b) 2D model with projective chiral symmetry along the $x$-direction (see equation~\eqref{eq:PrChiral_x}). (c) Dispersion relation of the SSH3 model. (d) 2D the Brillouin zone for model (b). (e) Dispersion relation of a ribbon of model (b) with $N_y=11$ sites in the vertical direction. (c,e) We took $t=t'=0.1$, $s=\lambda'=0.2$, $\lambda = s' = 0.3$, $u=u'=u''=0.25$ and $v=v'=v''=0.75$.}
	\label{Fig:Pr_chiral} 
\end{figure*}

The rather simple algebraic condition~\eqref{eq:PrChiral_x} has surpizingly rich consequences. Let us first discuss the 1D case. The representative example of SSH3 is shown in Fig.~\ref{Fig:Pr_chiral}-(a) (we will come back to this example later). The infinite system has sublattice symmetry (i.e. $\left\{\Gamma, \mathcal H \right\} = 0$), hence, using the commutation relation~\eqref{eq:PrChiral_x} we obtain the following identity satisfied by the Bloch Hamiltonian: 
\begin{equation} \label{eq:PrChiral_H}
\Gamma H(q) \Gamma = -H(q+\pi) , 
\end{equation} 
where $q$ is the adimensional Bloch wavenumber (the derivation of this identity is detailed in Appendix~\ref{App:PrH_derivation}). We introduce the two invariant points $X' = \pi/2$ and $X'' = -\pi/2$, which both have the property that $q + \pi = -q$. When combined with time reversal symmetry $H(q)^* = H(-q)$, the relation~\eqref{eq:PrChiral_H} becomes 
\begin{equation} \label{eq:1D_PHS}
\Gamma H(X') \Gamma = -H(X')^* , 
\end{equation} 
with the same identity at $X''$. Hence, at the invariant points the Hamiltonian has the Particle-Hole Symmetry (PHS) of class $D$ superconductors~\cite{Ryu2010,Budich2013,Chiu2016}. In particular, when the number of site in a unit cell is odd, this Hamiltonian has a zero-energy mode. This can be directly seen by taking the determinant of equation~\eqref{eq:1D_PHS}, which gives $\det(H(X')) \in i \mathbb R$, and the hermiticity of the Hamiltonian giving $\det(H(X')) \in \mathbb R$. This implies that the middle band necessarily crosses the zero energy level at $q = \pm \pi/2$, and this zero-energy crossing is symmetry protected. 

A simple one dimensional example with such symmetries is given by a variant of the Su-Schrieffer-Heeger (SSH) model with three sites per unit cell, called the SSH3 model~\cite{Su1981,Guo2015,MartinezAlvarez2019,Anastasiadis2022,Sougleridis2024}. The dispersion relation showed in Fig.~\ref{Fig:Pr_chiral}-(b). As expected, the middle band crosses $E=0$ for $q = X'$ and $q = X''$ irrespectively of the hopping coefficients. 

In 2D, projective chiral symmetry has even more interesting consequences. In particular, it can lead to a novel class of topologically protected edge modes. Without loss of generality~\cite{Xiao2024}, we can assume that the lattice has projective chiral symmetry along one direction only, that is, it satisfies equation~\eqref{eq:PrChiral_x} as well as $[\Gamma, \tau_y] = 0$. In other words, the role of sublattices $\mathcal A$ and $\mathcal B$ are swapped when making a single translation along $x$, but are kept unchanged along $y$. This is illustrated with a representative example in Fig.~\ref{Fig:Pr_chiral}-(c). At the level of the Bloch Hamiltonian, this leads to the identity:  
\begin{equation} \label{eq:PrChiral_H_2D}
\Gamma H(q_x,q_y) \Gamma = -H(q_x+\pi,q_y) . 
\end{equation} 
We now define two invariant lines $q_x = \pm \pi/2$ (see Fig.~\ref{Fig:Pr_chiral}-(d)) along which we obtain an effective 1D Hamiltonian with PHS, i.e. using time reversal symmetry equation~\eqref{eq:PrChiral_H_2D} becomes: 
\begin{equation}
	\Gamma H(\pi/2,q_y) \Gamma = - H(\pi/2,-q_y)^* . 
\end{equation}
One can then define a $\mathbb Z_2$ topological invariant~\cite{Kitaev2001,DeGottardi2013,Wakatsuki2014} $\eta$ as 
\begin{equation} \label{eq:TopoInv}
(-1)^\eta = \mathrm{sign}[\mathrm{Pf}(\Gamma H(X'))\mathrm{Pf}(\Gamma H(M'))] , 
\end{equation} 
where $\mathrm{Pf}$ denotes the Pfaffian, and $X'$ and $M'$ are the invariant points (see Fig.~\ref{Fig:Pr_chiral}-(d)). Notice that by time reversal invariance, the same topological invariant is obtained by using the other invariant line (i.e. using the invariant points $X''$, $M''$ in equation~\eqref{eq:TopoInv}), so the system is really characterized by the single invariant of equation~\eqref{eq:TopoInv}. A direct consequence of $\eta = 1 \; \mathrm{mod} \; 2$ is that if the system is gapped around $E=0$, it possesses topologically protected edge waves along any boundary parallel to $\tau_x$, according to the bulk-boundary correspondence~\cite{Kitaev2001,DeGottardi2013,Wakatsuki2014}. In addition, the edge modes have the same property as in 1D projective chiral models: its band crosses the zero-energy line at $q_x = \pm \pi/2$. This is shown in Fig.~\ref{Fig:Pr_chiral}-(e) where we show the modes of a finite width ribbon along the $x$-direction of the lattice of Fig.~\ref{Fig:Pr_chiral}-(c). It should be emphasized that these edge modes are protected solely by projective chiral symmetry. In particular, they require no mirror or rotation symmetry, as evidenced in the example of Fig.~\ref{Fig:Pr_chiral}-(c-e). 

Notice that while similar to~\cite{Xiao2024}, our topological classification differs from their due to the presence of time reversal invariant, which allows for a direct link to class $D$ Hamiltonians. A classification in higher dimensions is proposed in appendix~\ref{App:Topo_Class}.

%
%

\medskip
\noindent
\textit{Scattering on defects.}
Projective chiral symmetry also has exotic consequences when considering the scattering over defects. In particular, the obtained topological edge waves are generally not immune to backscattering. However, two consecutive identical defects will be completely invisible at $E=0$, as long as chiral symmetry is preserved. In other words, any chiral preserving defect can be made invisible by placing an exact replication right after. Note that this phenomenon is similar to ``anti-reflection'' structures~\cite{Horodynski2022}, but here the anti-reflection is obtained with a copy of the first defect. 

As discussed above, the relation~\eqref{eq:PrChiral_H} implies a zero energy crossing at $k=\pm \pi/2$. This condition can be expressed in terms of the transfer matrix $M$ across a single (projective chiral) cell~\cite{Dwivedi2016,Coutant2024,Torres2024}. Recalling that the eigenvalues of $M$ are $e^{\pm i q}$ in general, this means that at $E=0$, these eigenvalues are $\pm i$. In the case of a one-dimensional chain, the transfer matrix is $2\times2$ and $\pm i$ are the only eigenvalues, which implies that $\mathrm{Tr}(M) = 0$. Using time reversal symmetry $\mathrm{det}(M) = 1$, the Caley-Hamilton theorem gives: 
\begin{equation}
	M^2 = -\mathrm{I}_2 ,
\end{equation} 
with $\mathrm{I}_2$ the identity matrix. This condition guarantees that across a double defect an incoming mode, here the topological edge mode, will be fully transmitted. Additionally, the phase shift is locked to be $\pi$, independently of the defect. This means that the edge waves not only undergo perfect transmission but full invisibility at $E=0$. 

While the above arguments are exact in on-dimensional systems, they are only approximate for two-dimensional lattices. The reason is that the transfer matrix also includes evanescent modes, and hence is not $2\times2$ in general~\cite{Dwivedi2016,Coutant2024,Torres2024}. However, their contributions is generally small, and since there is only a single pair of propagating modes near $E=0$, we expect a near invisibility of double defect to edge waves in two dimensions. 

\begin{figure*}[htp]
	\centering
	\includegraphics[width=\textwidth]{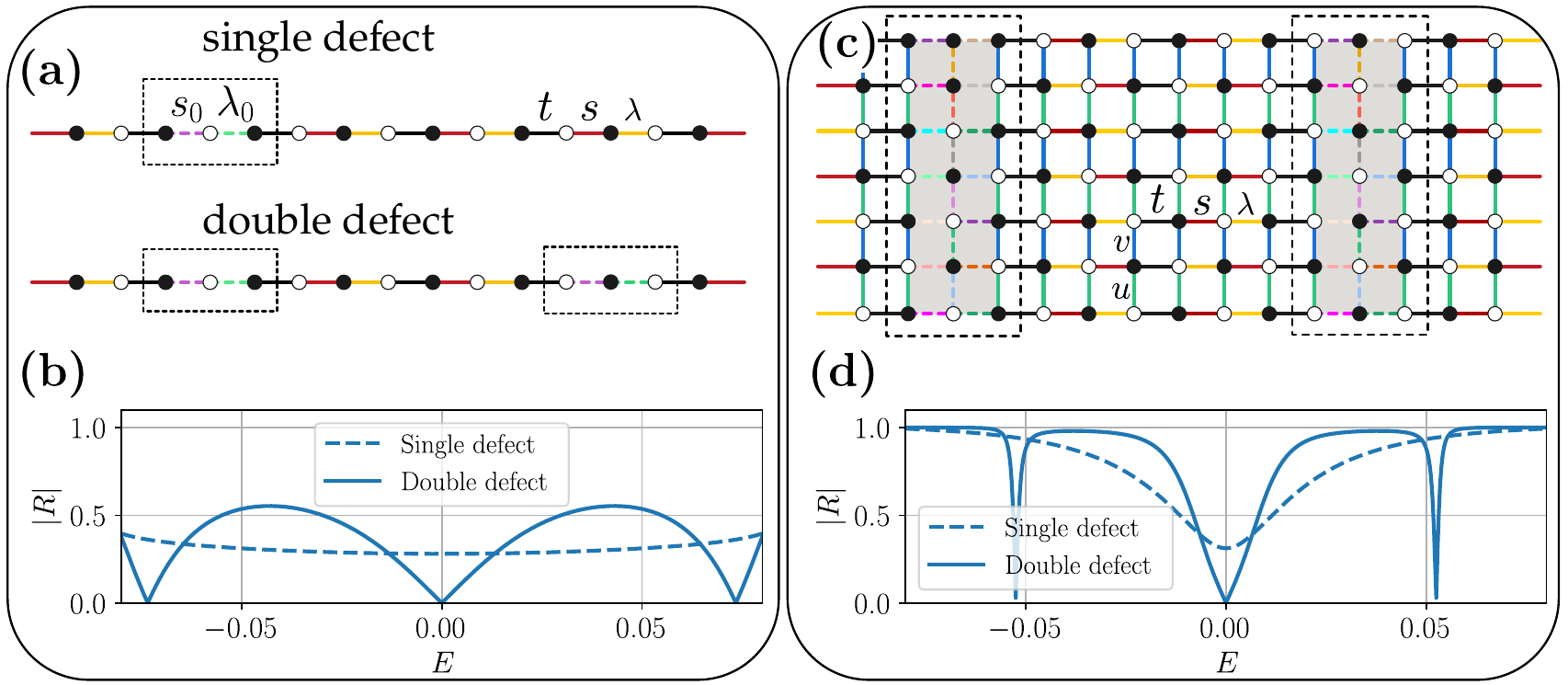} 
	\caption{(a) Scattering configurations in an SSH3 chain over a single or double defect. (b) Reflection coefficients for a single or double defect. In the chain the hopping coefficients are $s=0.2$, $\lambda = 0.3$ and $t=0.1$. For the defect we use $s_0 = 0.1$ and $\lambda_0 = 0.2$. (c-f) Scattering in a ribbon of the lattice in Fig.~\ref{Fig:Pr_chiral}-(c,d) with $u = u'= u'' = 0.1$, $v = v' = v'' = 0.6$, $t = t' = 0.3$, $s = \lambda' = 0.2$, $s' = \lambda = 0.1$. The width of the ribbon is 7 sites. 	(c) Configuration with a defect made of a slab of modified hopping (chosen randomly). The double defect is obtained by putting the exact same defect further. (d) Reflection coefficient for the single or double defects of (c). }
	\label{Fig:scattering}
\end{figure*}

This peculiar phenomena of symmetry protected invisibility is confirmed by a direct resolution of the scattering problem using a transfer matrix approach (similarly to what was done in~\cite{Coutant2024,Torres2024}). Moreover, we chose the ribbons to have an odd number of sites along the transverse direction. This allows us to have edge modes only on the lower edge, because the lower and upper edges correspond to different unit cells, and hence, have different topological invariants (see~\cite{Coutant2021a}). The results are shown in Fig.~\ref{Fig:scattering}. 

Fig.~\ref{Fig:scattering}-(a) and (b) show the SSH3 model of Fig.~\ref{Fig:Pr_chiral}-(a,b) containing a (single or double) defect made of two altered hopping coefficients. In the case of a single defect, the edge waves are significantly reflected for all energies considered ($30\% \lesssim |R| \lesssim 40\%$). When the defect is doubled, we see that at $E=0$ the reflection coefficient is exactly $0$ and it is significantly reduced for $E \approx 0$.

In Fig.~\ref{Fig:scattering}-(c,d), we show a ribbon of the two-dimensional lattice of Fig.~\ref{Fig:Pr_chiral}-(c,d). We consider a column of random hopping coefficients. Edge modes around $E=0$ are mostly reflected against the single defect. We then add an exact copy of the defect column further. In this latter case, edge modes are nearly perfectly transmitted at $E=0$. Note that for the double defect there exists other modes with energies $E\neq 0$ which are (nearly) perfectly transmitted, likely associated with resonances of the cavity formed by the double defect. However, such (nearly) perfect transmissions and their energies are not fixed, as opposed to the zero reflection at $E=0$ which is protected by the projective chiral symmetry.

\medskip
\noindent
\textit{-- Realization in acoustic networks.}
We now present an experimental observation of the topological edge modes. For this, we rely on the exact mapping between acoustic networks and lattice model Hamiltonian, that has been previously exploited in~\cite{Zheng2019,Zheng2020,Coutant2020a,Coutant2021,Coutant2021a,Coutant2024,Cote2026}. We quickly recall how this mapping works, and refer to previously cited works for details. An acoustic network is made of slender tubes all with the equal length $L$ but with possibly different cross sections connected as a graph. The slenderness (i.e. typical transverse length $L_\perp \ll L$) implies that the pressure inside each tube obeys the one-dimensional Helmholtz equation $p'' + k^2 p = 0$, with wavenumber $k = \omega/c_0$, $\omega$ the frequency and $c_0$ the speed of sound in air. $p$ is the complex stationary field associated with time dependent pressure $\Re(p e^{-i \omega t})$. At each vertex of the graph, the pressure is continuous, and conservation of acoustic flow rate leads to the so-called Kirchhoff condition~\cite{Kuchment2008,Delourme2017,Delourme2024}: 
\begin{equation}
	\sum_{m \in \mathcal{E}_n} \sigma_{nm} p'_{m}(n) = 0 , 
\end{equation}
where $\mathcal{E}_n$ is the set of vertices connected to a given vertex $n$, $\sigma_{nm}$ is the cross-section of the tube connecting vertices $n$ and $m$, $p'_{m}(n)$ is the derivatives pointed inward of the pressure at vertex $n$ coming from $m$. Because all tubes have the same length, the problem reduces to a generalized eigenvalue problem on the vector $\psi = (p(n))$ containing pressure values at each vertex: 
\begin{equation} \label{eq:GenEig_pressure}
	E \hat \mu \psi = \hat \sigma \psi . 
\end{equation}
In this equation, $\hat \sigma$ is the (symmetric) matrix containing the cross-section values $\sigma_{nm}$, $E$ is the eigenvalue related to the acoustic frequency by the relation $E = \cos(kL)$, and $\hat \mu$ is analogous to a mass matrix, which is diagonal with the component $n$ being equal to $\hat \mu_{nn} = \sum_{m \in \mathcal{E}_n} \sigma_{nm}$. This generalized eigenvalue problem can now be mapped to a lattice model by using the vector $\Psi = \hat \mu^{1/2} \psi$. Doing so we obtain the eigenvalue problem $E \Psi = \mathcal{H} \Psi$ with the Hamiltonian $\mathcal{H} = \hat \mu^{-1/2} \hat \sigma \hat \mu^{-1/2}$. The Hamiltonian matrix elements read: 
\begin{equation} \label{eq:Hlat_fromAc}
	\mathcal{H}_{nm} = \frac{\sigma_{nm}}{\sqrt{\sum_{m' \in \mathcal{E}_n} \sigma_{nm'} \sum_{m' \in \mathcal{E}_m} \sigma_{mm'}}} . 
\end{equation}
Notice from this equation that hoppings located on the edges of the network will take slightly different values from that in the bulk (see the discussion in~\cite{Coutant2021a} in the 1D context). Of course, this does not change the topological invariant, nor does it affect the presence or absence of edge modes. 

When the system is excited by a fixed frequency source located on vertex $j$, the pressure response at vertex $\ell$ is given by the Green function. To take into account sources of dissipation, we evaluate it at the complex energy $E+i\nu$, where $E = \cos(kL)$ as before and $\nu>0$ accounts for dissipation. Using~\ref{eq:GenEig_pressure}, this means we compute the function: 
\begin{equation} \label{eq:Gmat}
	G_{\ell j}(E) = [((E+i\nu)\hat \mu - \hat \sigma)^{-1}]_{\ell j} . 
\end{equation}
In the next section, we use $\nu = 0.02$, which represent fairly well the amount of dissipation (see appendix~\ref{App:Dissip} for details on the estimation of $\nu$). 

\begin{figure}[htp]
	\centering
	\includegraphics[width=\columnwidth]{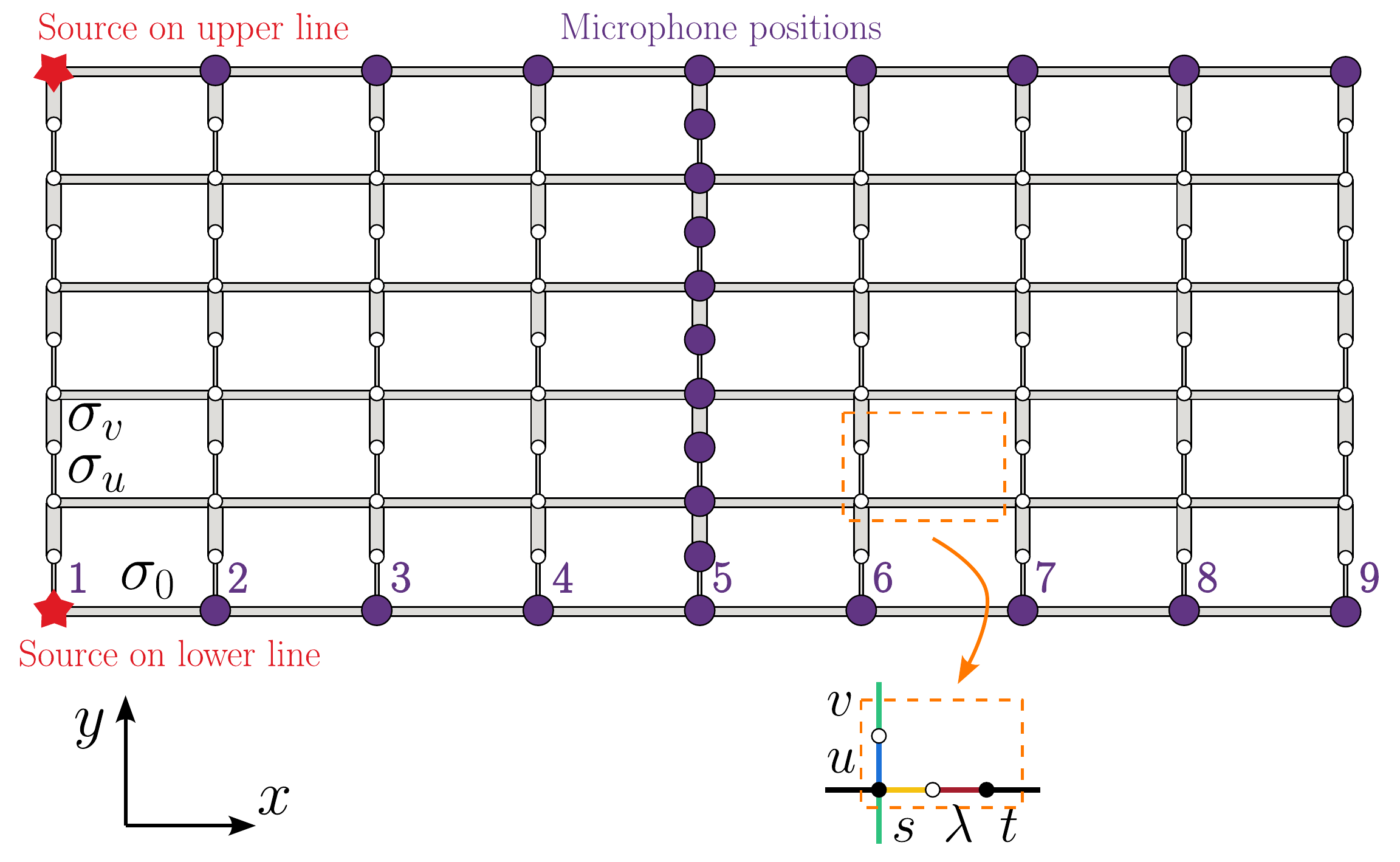} 
	\caption{Schematics of the experimental setup. We emphasize one unit cell and map it to a lattice cell. With the section ratio, we obtain that $u = \frac{1}{6\sqrt{5}}$, $v = \frac{3}{2\sqrt{5}}$, $s=\frac13$, $\lambda = \frac12$ and $t=\frac13$. Notice also that in the lattice, hopping coefficients on the boundaries have slightly different values due to different connectivity, as can be seen from equation~\ref{eq:Hlat_fromAc}. The two stars show the two positions of the source used in the experiment, and the purple circles the positions of the microphones (we also measure at the source positions). The numbers labels the columns.}
	\label{Fig:ExpSetup} 
\end{figure}

\begin{figure*}[htp]
	\centering
	\includegraphics[width=\textwidth]{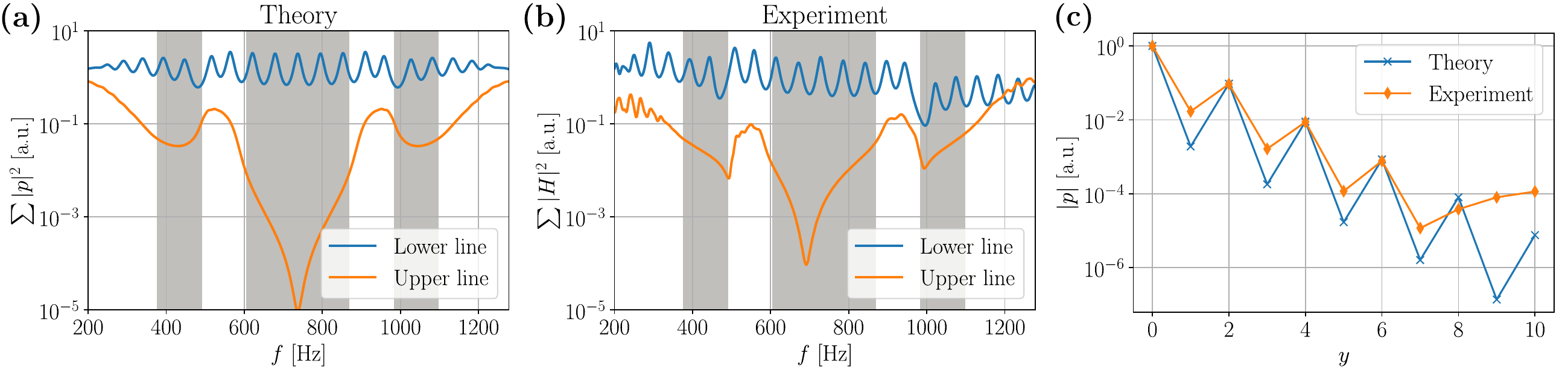} 
	\caption{(a,b) Frequency response functions $F_\alpha(f)$ on the upper and lower lines, with a source on the upper or lower line. (a) Theory derived from equation~\ref{eq:Gmat}. (b) Experimental results. The signal is normalized (multiplicative frequency independent constant) so as to minimize the square norm difference with the corresponding theoretical signal. (c) Response profile at $f_0$ along the $5^{\rm th}$ column (see Fig.~\ref{Fig:ExpSetup}). Both are normalized to be unity on the lower line.}
	\label{Fig:Results} 
\end{figure*}

\medskip
\noindent
\textit{-- Experimental observation of edge waves.}
We now use this correspondence to build an acoustic network mapped to a lattice model displaying topological edge modes protected by projective chiral symmetry. To ease the building the network, we chose a configuration with several hoppings set to 0 (compared to Fig.~\ref{Fig:Pr_chiral}-(c)), which allow us to use fewer tubes. As shown in Fig.~\ref{Fig:ExpSetup}, along the $y$-axis, we use tubes of length $L = 116 \; \mathrm{mm}$ and alternating sections $\sigma_u$ and $\sigma_v = 10.6\sigma_u$; along the $x$-axis, we use tubes of length $3L$ and section $\sigma_0 = 4\sigma_u$. Using the mapping of equation~\ref{eq:Hlat_fromAc} this network is equivalent to a lattice model, whose unit cell is shown in Fig.~\ref{Fig:ExpSetup}. The lattice is constructed to be incommensurable, that is, the edges on the upper part is associated with a different unit cell than the edge on the lower part (similarly to the SSH model, see e.g.~\cite{Coutant2021a}). As a result, we expect the lower edge to host edge modes, while the upper edge does not.

At each vertex, the tubes are connected with a 3D printed module that can host a microphone or a source. Microphones measure the pressure at a vertex with
a sensitivity of 24 mV/Pa. The source consist in a compression chamber (BMS 4552-8) placed at a given vertex. We then perform a synchronized frequency swipe from 50 Hz to 2000 Hz and use the method of~\cite{Novak2015} to extract each frequency response $h_j$ on each vertex $j$ obtained as the ratio of the measured pressure over the source input signal. To confirm the presence of edge modes, we performed the experiment twice: first with the source on the upper line and second with the source on the lower line. We then build a frequency response function by summing the square modulii of $h_j$ along the same line as the source (i.e. along $x$). In other words, we define two functions: 
\begin{equation}
	F_\alpha(f) = \sum_{j \in \alpha} |h_j(f)|^2 , 
\end{equation} 
where $\alpha$ represents the set of vertices on a given line. In Fig.~\ref{Fig:Results}-(a,b), we show the frequency response functions $F_\alpha$ when the sum is over the lower or upper line. In this figure, we also emphasize the gaps predicted by the discrete model in grey. We see very clearly that for the upper edge, propagation is strongly inhibited in gaps, while along the lower edge, propagation is observed for all frequencies. This is allowed by the presence of edge waves within the gaps. 

To confirm the localized character of edge modes, we show the frequency response functions $F_\alpha$ along the middle column ($5^{\rm th}$) at the frequency $f_0 \approx 739$ Hz in Fig.~\ref{Fig:Results}-(c). Our measurements exhibits a clear exponential decay, in excellent agreement with the theoretical model, which is a characteristic signature of localization. 

\medskip
\noindent
\textit{-- Conclusion} This work shows that the combination of chiral symmetry anti-commuting with translation in time reversal systems leads to rich consequences: new topological phases, protected band crossing at $E=0$ and backscattering immunity across double defects. These findings are confirmed by an experimental observation of topological edge modes in an acoustic network. Our experimental realization supports the fact that the projective chiral symmetry can be effectively leveraged in classical architectured materials to control wave propagation. 

\medskip
\noindent
\textit{-- Uses of generative artificial intelligence}
No generative artificial intelligence tools has been used at any stage of this work. AC is signatory of the {\it Atecopol Manifesto}~\cite{AtelierEcopol2025}.

%
%

\bibliography{/home/acoutant/Nextcloud_CNRS/Recherche/Notes/SuperBib.bib}

%
%

\newpage 
\appendix
\section{Details about identity~\ref{eq:PrChiral_H}}
\label{App:PrH_derivation}

We provide here a detailed derivation of the relation~\eqref{eq:PrChiral_H}. For this, we consider any solution $\Phi$ of the global (infinite) problem 
\begin{equation} \label{AppEq:Global_H}
	\mathcal H \Phi = E \Phi . 
\end{equation}
The Bloch theorem tells us that we can look for a basis of solutions of~\eqref{AppEq:Global_H} of the form 
\begin{equation} \label{eq:TensorBlochWave}
	\Phi = e^{i m q} \otimes \varphi , 
\end{equation}
with $\varphi$ a column vector of size corresponding to the number of sites inside a unit cell. We use this tensor product notation for its compactness, and recall that it means that the components inside unit cell number $m$ is: 
\begin{equation}
	[\Phi(q)]_m = e^{i m q} \varphi . 
\end{equation}
By construction of the Bloch Hamiltonian $H(q)$, we have: 
\begin{equation} \label{eq:TensorBlochHam}
	\mathcal H (e^{i m q} \otimes \varphi) = e^{i m q} \otimes (H(q) \varphi) . 
\end{equation}
To investigate the effect of chiral symmetry, we distinguish the global chiral operator $\tilde{\Gamma} = \mathrm{diag}(-\mathbf{1}_{\mathcal A},\mathbf{1}_{\mathcal B})$ acting on infinite sized vectors, and the local one $\Gamma = \mathrm{diag}(-\mathbf{1}_{\mathcal A},\mathbf{1}_{\mathcal B})$ acting on unit cell sized vectors. Notice that in the main text, for a smoother reading we have used the same symbol to denote both operations, but depending on whether it applies to an infinite or cell Hamiltonian we mean $\tilde{\Gamma}$ or $\Gamma$. Applying it on Bloch waves, we have: 
\begin{equation} \label{eq:TensorBlochChiral}
	\tilde{\Gamma} (e^{i m q} \otimes \varphi) = (-1)^m  e^{i m q} \otimes (\Gamma \varphi) , 
\end{equation}
where the phase $(-1)^m$ is a direct consequence of the anti-commutation relation $\{\tilde{\Gamma}, \tau_x\} = 0$. In other words, the role of this term is to swap the roles of sublattices $\mathcal{A}$ and $\mathcal{B}$ every other cell, as illustrated in Fig.\ref{Fig:Pr_chiral}-(a). Moreover, this term can be absorbed in a shift of Bloch wavenumber since $(-1)^m  e^{i m q} = e^{i m (q+\pi)} $. Now, since the global lattice has chiral symmetry, we have: 
\begin{equation}
	\tilde{\Gamma} \mathcal{H} \tilde{\Gamma} = - \mathcal{H} . 
\end{equation}
Applying this identity to a tensor product as in~\ref{eq:TensorBlochWave}, and using the identities~\ref{eq:TensorBlochHam} and~\ref{eq:TensorBlochChiral}, we obtain: 
\begin{equation}
	\Gamma H(q+\pi) \Gamma = -H(q) , 
\end{equation}
which, after using the $2\pi$-periodicity of the Hamiltonian, gives the identity~\ref{eq:PrChiral_H} used in the article.

\section{Topological classification of time reversal projective chiral models}
\label{App:Topo_Class}
We quickly provide the general topological classification of time reversal invariant Hamiltonians with projective chiral symmetry. In $d$ dimension, we write the coordinates $(x_1,\dots x_d)$, and the corresponding momentum coordinates $(q_1,\dots q_d)$. As in the main text, we assume that the chiral operation anti-commutes with $\tau_1$, and commutes with all other translations. Along the hypersurface $q_1 = \pi/2$ in the Brillouin zone, the Bloch Hamiltonian satisfies: 
\begin{equation}
	\Gamma H\left(\frac{\pi}{2}, \mathbf{q}\right) \Gamma = - H \left(\frac{\pi}{2},-\mathbf{q}\right)^* , 
\end{equation}
where $\mathbf{q}$ is the $d-1$-momentum vector $(q_2,\dots q_d)$. Hence, we see that the topological classification reduces to that of type $D$ Hamiltonian in dimension $d-1$~\cite{Ryu2010,Teo2010}. Notice that the hypersurface $q_1 = -\pi/2$ leads to the same relation and invariant. We deduce the results of Table~\ref{Table:Topo_Class}. We notice important differences with the classification without time reversal invariance, as proposed in~\cite{Xiao2024}. For instance $d=3$ is trivial in non time reversal invariant classes, but leads to a $\mathbb{Z}$ class here. 

\begin{table}[!htp]
	\centering 
	\begin{tabular}{|c|ccccccccc|}
		\hline 
		& dim. & $1$ & $2$ & $3$ & $4$ & $5$ & $6$ & $7$ & $8$ \\
		Sym. &  & \hspace{15pt} & \hspace{15pt} & \hspace{15pt} & \hspace{15pt} & \hspace{15pt} & \hspace{15pt} & \hspace{15pt} & \\
		\hline
		$\{\Gamma, \tau_x\}$ & & $\mathbb Z_2$ & $\mathbb Z_2$ & $\mathbb Z$ & $0$ & $0$ & $0$ & $2\mathbb{Z}$ & $0$ \\
		\hline
	\end{tabular}
	\caption{Topological classification of time reversal invariant Hamiltonians with projective chiral symmetry. As for real class Hamiltonians~\cite{Ryu2010,Teo2010}, the classification is $8$-periodic in dimension due to Bott periodicity.}
	\label{Table:Topo_Class}
\end{table}

\section{A separable two-dimensional model}
\begin{figure}[!h]
	\centering
	\includegraphics[width=0.7\columnwidth]{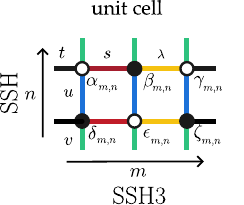}
	\caption{Unit cell of the SSH$\otimes$SSH3.}
	\label{Fig:SSHxSSH3} 
\end{figure}

We detail here a the properties of a simpler model exhibiting extra symmetries in addition to the projective chiral symmetry discussed in the main text. The system consists of a square lattice constructed from the standard SSH chain in the vertical direction and the SSH3 chain in the horizontal direction, see Fig.~\ref{Fig:SSHxSSH3}. 
The discrete eigenvalue problem for this system is
\begin{eqnarray*}
	E \alpha_{m,n} &=& t\gamma_{m-1,n} + v\delta_{m,n+1} + u \delta_{m,n} + s\beta_{m,n},\\
	E \beta_{m,n} &=& s\alpha_{m,n} + v\epsilon_{m,n+1} + u \epsilon_{m,n} + \lambda \gamma_{m,n}, \\
	E \gamma_{m,n} &=& t\alpha_{m+1,n} + \lambda \beta_{m,n} + v\zeta_{m,n+1} + u \zeta_{m,n} , \\
	E \delta_{m,n} &=& t\zeta_{m-1,n} + s\epsilon_{m,n} + v\alpha_{m,n-1} + u \alpha_{m,n}, \\
	E \epsilon_{m,n} &=& s\delta_{m,n} + v\beta_{m,n-1} + u \beta_{m,n} + \lambda \zeta_{m,n}, \\ 
	E \zeta_{m,n} &=& t\delta_{m+1,n} + \lambda \epsilon_{m,n} + v\gamma_{m,n-1} + u \gamma_{m,n}.
\end{eqnarray*}
The above model is separable~\cite{Cerjan2020,Zhu2020,Benalcazar2020,Coutant2020a,Coutant2021} and we can look for solutions of the form
\begin{equation}
	\begin{pmatrix}
		\alpha_{m,n} \\
		\beta_{m,n} \\
		\gamma_{m,n} \\
		\delta_{m,n} \\
		\epsilon_{m,n} \\
		\zeta_{m,n}
	\end{pmatrix} = \begin{pmatrix}
		\psi_\alpha^m \phi_\alpha^n \\
		\psi_\beta^m \phi_\alpha^n \\
		\psi_\gamma^m \phi_\alpha^n \\
		\psi_\alpha^m \phi_\beta^n \\
		\psi_\beta^m \phi_\beta^n \\
		\psi_\gamma^m \phi_\beta^n
	\end{pmatrix} \dot{=} \ \psi^m \otimes \phi^n
\end{equation}
The left hand side is a solution to the eigenvalue problem if the vectors $\psi^m$ and $\phi^n$
simultaneously satisfy
\begin{eqnarray}
	E_x \psi_\alpha^m &=& t\psi_\gamma^{m-1} + s\psi_\beta^m,\\
	E_x \psi_\beta^m &=& s\psi_\alpha^{m} + \lambda\psi_\gamma^m,\\
	E_x \psi_\gamma^m &=& t\psi_\alpha^{m+1} + \lambda\psi_\beta^m,
\end{eqnarray}
and
\begin{eqnarray}
	E_y \phi_\alpha^m &=& v\phi_\beta^{n+1} + u\phi_\beta^n,\\
	E_y \phi_\beta^m &=& v\phi_\alpha^{n-1} + u\phi_\alpha^n,
\end{eqnarray}
with $E = E_x + E_y$
In other words, $\phi^m$ and $\psi^n$ are respectively solutions of the 1D SSH3 and SSH chains. 
\subsection{Infinite network}
In an infinite network we can further look for Bloch wave solutions
\begin{equation}
	\psi^m \otimes \phi^n = e^{imq_x + inq_y} \psi \otimes \phi.
\end{equation}
Using the separability, the dispersion relation of the 2D network can be obtained from the one-dimensional subsystems
\begin{equation}
	E(q_x,q_y) = E_x(q_x) + E_y(q_y).
\end{equation}
More explicitly, $E_x$ corresponds to the energy eigenvalues of the SSH3 chain and $E_y$ to the ones of the usual SSH chain
\begin{eqnarray}
	E_y(q_y) &=& \pm|u+ve^{iq_y}|.
\end{eqnarray}
$E_x$ are the solutions of the dispersion relation
\begin{eqnarray}
	E_x^3 - \left( s^2+\lambda^2 + t^2\right)E_x - 2s\lambda t {\rm{cos}}(q_x) =0.
\end{eqnarray}
The separability further implies that the bulk Hamiltonian can be decomposed as
\begin{equation}
	H(q_x,q_y) = I_2\otimes H_x + H_y \otimes I_3,
\end{equation}
where $H_{x/y}$ are the bulk Hamiltonian of the SSH3/SSH chain respectively and $I_N$ is the $N\times N$ identity matrix.
This allows us to compute the Pfaffian involved in the construction of the topological invariant explicitly
\begin{eqnarray*}
	{\rm Pf}[\Gamma H(X')] &=& -(u + v) (s^2 + t^2 + \lambda^2 + (u + v)^2), \\
	{\rm Pf}[\Gamma H(M')] &=& -(u-v) (s^2 + t^2 + \lambda^2 - (u - v)^2 ).
\end{eqnarray*}
Knowing the values of the hopping coefficient, it is therefore straightforward to compute the topological invariant and deduce the topological phase of the system. Once the topological phase is known, one can then modify the values of the hopping parameters and break the separability without affecting the topological invariant provided that the gap remains open.

\subsection{Ribbon configuration}
\begin{figure*}[!htp]
	\centering
	\includegraphics[width=\textwidth]{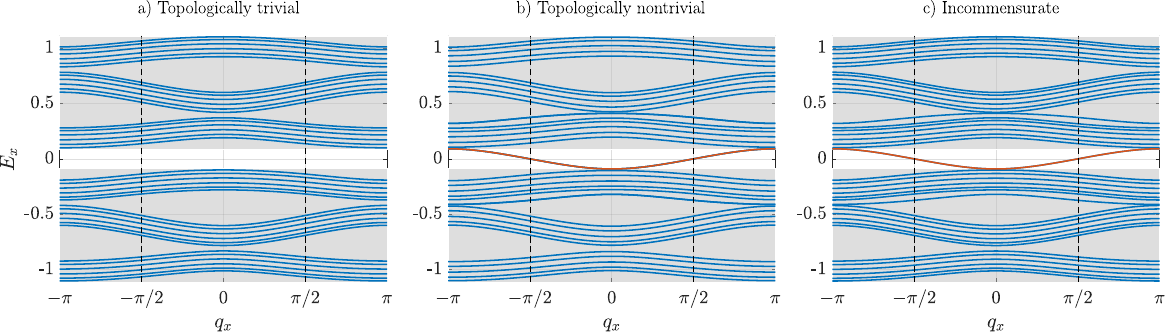}
	\caption{Dispersion relation of separable ribbon with $s=0.1$, $\lambda=0.2$, $t=0.3$, $(u,v)=(0.1.0.6)$ in the topologically trivial phase (panel a)) and  $(u,v)=(0.6.0.1)$ in the topologically nontrivial phase (panel b)). The lightgrey background represent the energy bands of the bulk. We can see in both the topologically nontrivial phase and the incommensurate ribbon the existence of an edge mode band located completely within the bulk gap (dark orange curve).}
	\label{Fig:ribbon} 
\end{figure*}

The separability properties also applies to the case of the ribbon configuration. The energy eigenvalues are given by
\begin{equation}
	E = E_x + E_y^{(j)},
\end{equation}
where $E_y^{(j)}$ are the energies of the finite SSH chain in the $y$ direction. Therefore, the band structure is composed of copies of the band diagram of the SSH3 chain shifted by the energies of the finite SSH chain. In particular, if the ribbon is incommensurate in the vertical direction, it will admit an edge mode with $E_y^{(j)} = 0$ and therefore its band will cross the zero-energy line at $q_x = \pi/2$. Provided that the gap of the vertical SSH is sufficiently large, its band will be entirely in the energy gap of the bulk. This is illustrated in Fig.~\ref{Fig:ribbon}.

\section{Acoustic dissipation model}
\label{App:Dissip}

There are several identified sources of dissipation in the proposed experiment. The main ones are possibles leaks at tube connections and thermo-viscous dissipation inside the tubes. The first source is minimized by carefully adjusting the sizes of 3D-printed modules and sealing with thread seal tape. The second source can be estimated by using a complex valued wave number $\underline{k}$ as a function of the frequency. When the tube radius is much larger than the viscous boundary layer, we have that 
\begin{equation} \label{eq:DissipRel}
	\underline{k}(f) = \frac{2\pi f}{c_0} + (1-i)\zeta \frac{\sqrt{f}}{R_0} , 
\end{equation}
where $R_0$ is the radius of the considered tube, and $\zeta$ depends on temperature thermodynamic properties of the medium. In air at $T = 20^\circ$C, we have $\zeta = 3.10^{-5} \mathrm{s}^{1/2}$ (see for instance p.~243 of~\cite{Chaigne2016} for details). It is tempting to simply use the same mapping between acoustic networks and lattice Hamiltonian with the complex energy $E = \cos(\underline{k} L)$. The problem is that $\underline{k}$ is not the same for all tubes since it depends on the tube radius. To avoid this issue, we use a simplified dissipation model. First, we neglect the real part of the correction in~\eqref{eq:DissipRel} (dispersive correction). We then expend $E = \cos(\underline{k} L)$ at first order in the dissipative correction, and evaluate it at the main frequency of interest, that is $f_0$. We find that $E = \cos(kL) + i \nu$ with $\nu \simeq 0.02$.

\section{More on the experimental results}
\label{App:More_Exp}

When an acoustic waves propagates across a cross-section discontinuity in a tube, departures from the 1D approximation can be encoded in an effective correction of its length, the so-called ``added length''~\cite{Jaouen2018}. In this work, to minimize these higher dimensional effects, we used slightly different tube length, so that the effective length $L = 116 \mathrm{mm}$ is the same for all tubes. For this, when a small tube of radius $r_s$ is connected to a larger tube of radius $r_b$, its physical length $L_{\rm phys}$ is taken to be: 
\begin{equation}
	L_{\rm phys} = L - 0.85 r_s \left(1-1.33 \frac{r_s}{r_b}\right) . 
\end{equation}
In the experimental setup reported in this work, we have used the physical values reported in Table~\ref{Table:Tubes}. 

\begin{table}[!htp]
	\centering 
		\begin{tabular}{|c|c|c|}
		\hline 
		& Length (mm) & Section radius (mm) \\
		\hline
		Tube $\sigma_u$ & 99.1 & 4 \\ 
		\hline
		Tube $\sigma_v$ & 105.1 & 13 \\ 
		\hline
		Tube $\sigma_0$ & 315.3 & 8 \\ 
		\hline 
	\end{tabular}
	\caption{Physical lengths and radii of the tubes. Each tube is named after its cross-section as used in the main text.}
	\label{Table:Tubes}
\end{table}

Taking into account the extra length inside the connecting modules, the effective length seen by acoustics waves is $L = 116 \mathrm{mm}$ for short tubes and $3L$ for long ones.

\begin{figure*}[!h]
	\centering
	\includegraphics[width=\textwidth]{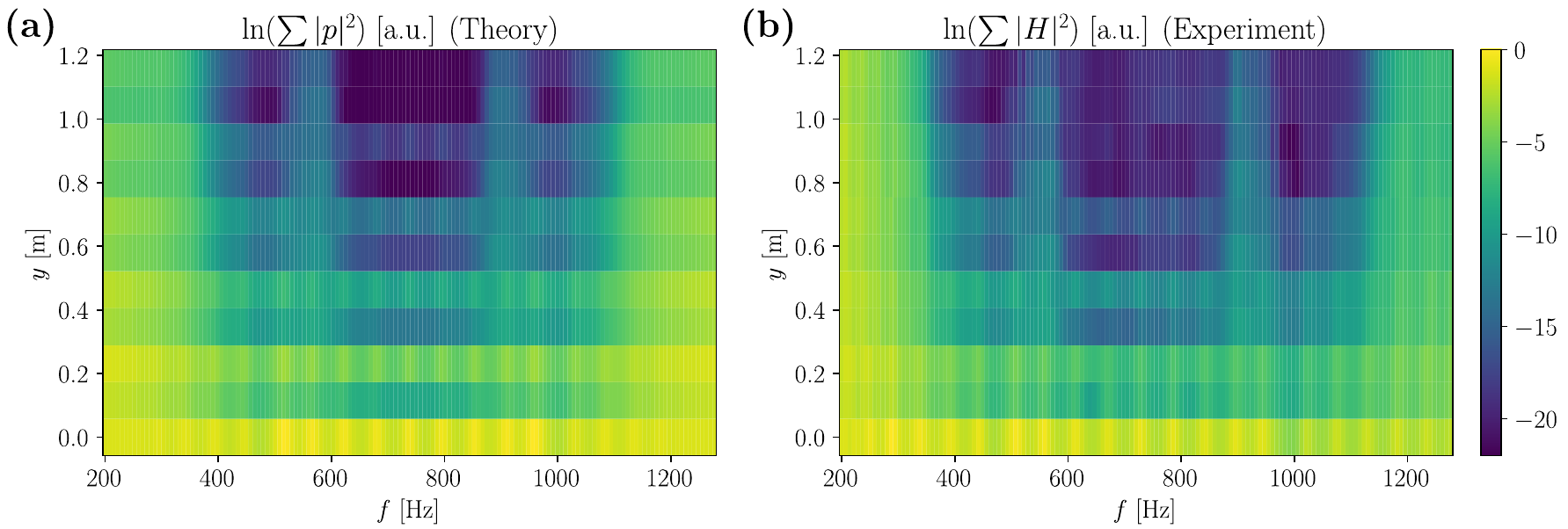} 
	\caption{(a,b) Frequency response functions $F_\alpha(f)$ as a function of frequency and vertical coordinate. (a) Theory derived from equation~\ref{eq:Gmat}. (b) Experimental results.}
	\label{Fig:2Dmaps} 
\end{figure*}

To complement the results of the main text, in Fig.~\ref{Fig:2Dmaps}, we show the frequency response functions $F_\alpha$ obtained by summing over each line when the source is in the lower line for different frequencies. We clearly see that in-gaps, the waves only travel along the lower edge, and is exponentially decreasing in the vertical direction.

\end{document}